\documentclass[useAMS,usenatbib]{mn2e}
                                                                                                                             
 \usepackage{epsfig}
 
\title[Limit on UHE Neutrino Flux from the Parkes Lunar Radio Cherenkov Experiment]{Limit on UHE Neutrino Flux from the  Parkes Lunar Radio Cherenkov Experiment}

\author[C. W. James et al.]{C. W. James$^{1}$\thanks{E-mail: clancy.james@adelaide.edu.au
(CWJ)}, R. M. Crocker$^{1}$,  R. D. Ekers$^{2}$, T. H. Hankins$^{3}$, J. D. O'Sullivan$^{2}$, \newauthor R. J. Protheroe$^{1}$\\
$^{1}$Department of Physics, School of Chemistry \& Physics, University of Adelaide, Adelaide SA 5000, Australia\\
$^{2}$Australia Telescope National Facility, PO Box 76, Epping NSW 1710, Australia\\
$^{3}$Physics Department, New Mexico Tech, Socorro, NM  87801, USA
}
\begin{document}
 
 
\pagerange{\pageref{firstpage}--\pageref{lastpage}} \pubyear{2006}
 
\maketitle
 
\label{firstpage}
 
\begin{abstract}
The first search for ultra-high energy (UHE) neutrinos using a
radio telescope was conducted by \citet{parkes}.  This was a
search for nanosecond duration radio Cherenkov pulses from ultra-high energy (UHE)
neutrino interactions in the lunar regolith, and was made using a
broad-bandwidth receiver fitted to the Parkes radio telescope,
Australia. At the time, no simulations were available to calculate
the experimental sensitivity and hence convert the null result into
a neutrino flux limit.

Proposed future experiments include the use
of broad-bandwidth receivers, making the sensitivity achieved by
the Parkes experiment highly relevant to the future prospects of
this field. We have therefore calculated the effective aperture
for the Parkes experiment and found that when pointing at the
lunar limb, the effective aperture at all neutrino energies was
superior to single-antenna, narrow-bandwidth experiments, and
that the detection threshold was comparable to that of the
double-antenna experiment at Goldstone. However, because only a
small fraction of the observing time was spent pointing
the limb, the Parkes experiment places only comparatively weak
limits on the UHE neutrino flux.  Future efforts should use multiple
telescopes and broad-bandwidth receivers.

\end{abstract}

 
\section{Introduction}

The properties expected of ultra-high energy (UHE) neutrinos make
them attractive targets for probing the high-energy
universe. Unlike the highest energy photons and cosmic rays, a
flux of neutrinos will not seriously suffer attenuation either at
its source or during propagation. Being uncharged, the paths of
neutrinos will remain unbent by galactic and intergalactic
magnetic fields, so any detected neutrino should point back to
its source. Additionally, any UHE neutrino flux should be very
sensitive to the nature and evolution of the sources of the
highest energy cosmic rays, providing a powerful discriminant
between models of UHE cosmic ray production \citep{stanev}.

The Lunar Cherenkov technique is a method by which UHE neutrinos
may in principle be detected. \citet{askaryan} described how a particle
cascade in a dense medium produces coherent Cherenkov radiation.
  If the medium is transparent at radio
frequencies, the radiation can escape and be detected remotely as
a narrow pulse of a few nanoseconds duration, corresponding to decimetre and greater wavelengths. The lunar regolith
(the outer layer of pulverised rock on the Moon's surface) is such a radio-transparent medium,
and as suggested by \citet{dag}, observations of the Moon with
ground-based radio-telescopes can be used to search for cascades
produced by UHE neutrino interactions. This technique works in
principle for both UHE cosmic rays and neutrinos, although
formation-zone effects are expected to significantly reduce the
cosmic ray signature \citep{radhep}.  At
UHE the neutrino-nucleon cross section is such that neutrinos
traversing the lunar diameter are severely attenuated.  Together
with subsequent shower and Cherenkov emission geometry, and
refraction at the lunar surface, this causes GHz-regime Cherenkov 
signals to appear to originate almost entirely from the limb of the Moon.

The first attempt to use the lunar regolith in the search for UHE
neutrinos was made at Parkes, Australia by \citet{parkes}. The 64-m
Parkes radio telescope was used to observe the Moon for approximately
10.5 hours using a wide-bandwidth dual-polarisation receiver. No real
events could be identified.  Subsequently, two independent experiments
utilising the technique also recorded null results,
the first by \citet{goldstone} being the Goldstone Lunar
Ultra-High Energy Neutrino Experiment (GLUE) that ran from 2001-2003 at
NASA's Goldstone Deep Space Communications Complex, USA, and the
second by \citet{kalyazin} conducted from 2002 to 2004 at the Kalyazin Radio
Astronomical Observatory, Russia. Importantly, both groups developed detailed simulations
of the technique (see, respectively, \citet{williams} and
\citet{beresnyak}). These were used to place limits on the UHE
neutrino flux, with the published GLUE limit producing severe
constraints on Z-burst UHE neutrino production models
\citep{goldstone}.

The search for UHE neutrinos now encompasses a wide range of
experiments, including the ANITA balloon experiment
\citep{anita}, low-frequency lunar observations with Westerbork
and LOFAR \citep{scholten}, and the Pierre Auger air-shower array
\citep{auger}. Though no UHE neutrinos have so far been detected,
limits placed on the UHE neutrino flux from various experiments, in particular the ANITA-lite limit \citep{anitalim}, have already ruled
out the more optimistic Z-burst models, and severely constrained the
remainder.

To ensure continuing competitiveness with these other efforts,
future Lunar Cherenkov observations should aim to utilise `next 
generation' radio-telescopes, in particular those designed as
large arrays of smaller stations with broad-bandwidth receivers
such as the planned SKA (Square Kilometre Array; \citet{SKA}), as discussed by \citet{prospects}. In order to improve real-time discrimination of Cherenkov
pulses from background noise and terrestrial radio-frequency interference (RFI), the full capabilities offered of such instruments in nano-second pulse detection will have to be exploited. This will require the latest in signal
processing technology. In parallel,
sophisticated simulations should be used to optimise observation
parameters such as frequency, beam pointing position, and bandwidth.
A first step in this process is an analysis of the broad-bandwidth
techniques developed at Parkes, the effectiveness of which we present here.

\section{Parkes Experiment}
\label{experiment}

The experiment at Parkes is described more fully by
\citet{parkesradhep}. Observations were on the nights of
16$^{\rm{th}}$, 17$^{\rm{th}}$, and 18$^{\rm{th}}$ of January,
1995. At the time of observation, the significant
limb-brightening effect had not been predicted and,
unfortunately, only 2 hours out of the total 10.5 hours of
observation time were spent pointing at the lunar limb. The remaining
8.5 hours, spent pointing at the centre, are not expected to
contribute significantly to the sensitivity, as the entire limb
was then outside the FWHM of the Parkes beam ($13'$ at the 
central frequency of 1.5 GHz).

In the experiment, data from two polarisation channels (LCP and
RCP) for a 500-MHz band centred at 1.425 GHz were
recorded. Triggering required a coincidence between two 100-MHz
bandwidth sub-bands, centred at 1.325 and 1.525 GHz, extracted
from either the LCP or RCP channel. The ionospheric delay between these sub-bands
(estimated at 10 ns) was corrected for by artificially delaying
the 1.525 GHz sub-band. Triggering occurred when the individual voltages in both the
sub-bands simultaneously exceeded an $8 \sigma$ level (eight times the measured
standard deviation of the oscilloscope voltage) for between 7.5 and 20 ns. 
This produced a trigger event approximately every two minutes.

The recorded data were processed to remove dispersion within each
band. This also helped filter out terrestrial interference,
which experiences no ionospheric delay (except any signals bounced off
the Moon, which experience twice the dispersion). Pulses of coherent
Cherenkov radiation were expected to be both 100\% linearly
polarised (and thus be received equally in both the LCP and RCP
channels) and broad-band; these properties have since been
verified in a series of experiments
\citep{slac, slac2, time_domain, slac3}. The recorded
data enabled candidate events to be tested for all these criteria,
assuming a dispersion in the range of zero to twice that expected,
a process which eliminated
all of the $\sim$700 triggered events. Thus it was concluded that
no Cherenkov pulse had been observed.

Only two 100MHz sub-bands could be used to form the trigger due to
the limitations of signal processing technology in the mid '90s,
where, ideally, the full 500MHz bandwidth with dual polarisation
would have been used. This proved to be the limiting sensitivity
as the remaining data (not used in triggering) proved more than
adequate for discriminating RFI and thermal fluctuations. To
demonstrate the usefulness of improved technology, it is useful to
speculate about what the Parkes sensitivity might have been had the entirety of
both data streams been de-dispersed in real-time and used to form
a trigger. A proper estimate requires a knowledge of the precise 
effect of dedispersion on the amplitudes of RFI, which was
responsible for the observed trigger rate and the setting of the
$8 \sigma$ level. A conservative estimate, however, for the sensitivity can
be obtained by assuming an identical trigger rate due to RFI,
knowing that de-dispersion will act to reduce the amplitudes of
the (undispersed) RFI signals. Therefore, we also present results
for an otherwise identical Parkes experiment in which a signal
strength of $8 \sigma$ in both of the full 500 MHz bands is
required for detection. It should be possible to reduce this down
to a $\sim 6 \sigma$ level on each channel in coincidence, which
is the requirement to eliminate events from normally distributed
thermal noise with 99.98\% confidence at 1 GHz sampling over a
10.5 h period.

\section{Simulations}
A Monte Carlo program was created to simulate the interactions of
UHE neutrinos with the Moon, the production and propagation of
coherent Cherenkov radiation, and the reception and triggering of
the signal by the Parkes antenna. The program instantiates similar physics to
the programs developed for the GLUE and the Kalyazin experiments.
For UHE neutrinos at discrete energies a lunar impact parameter, $r$, was
sampled from $p(r) \propto r$ for $0<r<r_m$ where $r_m$ is the
lunar radius.  The proportion of neutrinos detected by the
simulated experiment was recorded and used to give an estimate of the
detection probability per incident neutrino as a function of
neutrino energy. To estimate the effective experimental aperture,
the detection probability was multiplied by the physical lunar
aperture ($4 \pi^2 r_m^2$, $\approx 1.21 \times 10^8$ km$^2$-sr).

Both charged-current (CC) and neutral-current (NC) interactions
of UHE neutrinos were modelled, with energy-dependent
cross-sections taken from \citet{gandhixsections}. These interactions 
may initiate two kinds of showers. Electromagnetic showers consist
entirely of $\gamma$ and $e^{\pm}$, and are initiated only by the 
$e^{-}$/$e^{+}$ produced in a $\nu_{e}$/$\bar{\nu}_e$ CC-interaction,
(bremsstrahlung photons from the $\mu$/$\tau$ produced in
$\nu_{\mu}/\nu_{\tau}$ CC-interactions will be of insufficient
energy to begin detectable cascades). Hadronic
showers develop from both CC and NC interaction and consist of a
hadronic core surrounded by an electromagnetic component (see
\citet{AMZmethods} for a discussion of the relationship between
Cherenkov radiation and shower phenomenology). The
interaction inelasticity, $y$ (fraction of neutrino energy given to
hadronic showers), was sampled from the distributions used in
\citet{beresnyak}.  In the case of  $\nu_e$/$\bar{\nu}_e$ CC
interactions, where both electromagnetic and hadronic showers are present,
only the shower with the strongest emission at a given angle to
the shower axis was taken into account, because the relative
phase between the two components is unknown.  Neutrinos and anti-neutrinos were treated
identically, as were $\nu_{\mu}$ and $\nu_{\tau}$. One in three
incident neutrinos was assumed to be a $\nu_{e}$/$\bar{\nu}_e$ as we expect
complete flavour mixing during oscillation over extragalactic
distance scales \cite{flavourmixing}.

The lunar density was modelled with five distinct density layers,
with the densities of the inner four normalised so as to produce
the correct lunar mass as in \citet{williams}. The outer shell
-- nominally the regolith -- was modelled with 10m depth, for
consistency with both the simulations used by \citet{goldstone}
and the results of radar and optical studies discussed by
\citet{shkuratov}. A density of $1.8$ g/cm$^3$, and refractive
index $n=1.73$, was used for consistency with the Cherenkov
parameterisations of \citet{AMZlatest}.

Of these layers, only the regolith was treated as a suitable
medium for the production of coherent Cherenkov radiation
because of its known low attenuation at radio-frequencies. It
appears reasonable to assume that the mega-regolith -- a
layer of ejecta blankets between the regolith and underlying
bedrock in the lunar highlands, distinguished from the regolith
as outlined by \citet{short}, with an expected mean depth of
$\sim 2$ kms \citep{aggarwaloberbeck} -- may also exhibit low
radio-attenuation properties.  This region is treated by
\citet{scholten} as an extended regolith down to 500m depth.
Detailed modelling of the production of Cherenkov radiation and
radio-transmission through these surface layers of the Moon,
including the depth dependence of their electromagnetic
properties, is left to a future paper.

Neutrino interaction points were considered as point sources of
Cherenkov radiation, with the Cherenkov cone axis being in
the direction of the incoming neutrino. Previous simulations
\citep{williams, beresnyak} parameterized this radiation
according to the results of \citet{AMZEM} and \citet{AMZmethods}
in ice and scaled these results to the regolith according to the
prescription of \citet{AMZ_scaling}. Recently, \citet{AMZlatest}
obtained results for purely electromagnetic showers by simulating
the regolith directly, using a refractive index of $n=1.73$,
density $\rho=1.8$ g/cm$^3$, radiation length $X_0=22.59$
g/cm$^2$, and critical energy $E_C=40.0$ MeV (below which
ionisation losses dominate bremsstrahlung). The value of the
field strength at the Cherenkov angle was fitted as:
\begin{equation}
\label{peak}
R \, |E_{\theta=\theta_C}(\nu)| \, = \, 8.45 \times 10^{-8} E_s\,
\frac{\nu}{1+\left( \frac{\nu}{\nu_R} \right)^{\alpha} } ~~(\rm{V/MHz})
\end{equation}

for shower energy $E_s$ (TeV), frequency $\nu$ (GHz), and observation
distance (i.e.\ Earth-Moon distance) $R$ (m). The decoherence frequency
$\nu_R = 2.32$ GHz, and the scaling parameter $\alpha=1.32$, have both
been revised by \citet{AMZlatest} from their former values of $2.5-3.0$
GHz and $1.44$ respectively. Notably, the normalisation of
$8.45 \times 10^{-8}$ is approximately $30$\% lower than the simple
scaling relationships \citep{AMZ_scaling} would suggest, implying that
the GLUE and Kalyazin apertures were over-estimated, particularly to
lower neutrino energies (as discussed in Sec.\ \ref{results}).

Away from the Cherenkov angle, \citet{AMZlatest} found the simple
scaling relationships to be adequate to model the
decoherence. However, these authors note the new parameterisation
for radiation far from the Cherenkov angle is unreliable for
shower energies at which the Landau-Pomeranchuk-Migdal (LPM)
effect becomes important. This effect suppresses both the
bremsstrahlung and pair-production cross-sections when the
characteristic length of the interaction becomes comparable to 
distance between scattering centres (the atoms in the medium),
and is important at particle energies
above the LPM energy, $E_{\rm{LPM}}$. For the regolith,
$E_{\rm{LPM}}$$\approx$770 TeV, which covers the entire UHE range,
and therefore this new parameterisation
for angular spread will be inappropriate for cascades initiated by
UHE neutrinos.  Simulations are currently in progress with showers
in regolith for shower energies above the LPM energy.  In the
meantime, we have used Eqn.\ \ref{peak} to describe the peak
field strength at the Cherenkov angle for electromagnetic
showers, and take the angular dependence of purely
electromagnetic showers from \citet{AMZEM}, with the addition of
the $\sin \theta / \sin \theta_C$ term from \citet{AMZlatest}. The
characteristic width $\Delta \theta$ we assumed to scale with
$\rho/X_0/\sqrt{n^2-1}$ ($X_0$ the radiation length; 
$X_0 = 22.59$ g/cm$^2$ here). Thus we assume the Cherenkov cone to
be approximately 3.4 times as wide in the regolith as in ice.

The Cherenkov radiation from hadronic showers is derived almost
entirely from electromagnetic sub-showers resulting from
$\pi^0$-decay into $\gamma$-rays. Because of the similar
phenomenology, the peak pulse strength for hadronic showers can
be derived by multiplying the purely electromagnetic result,
Eqn.~\ref{peak}, by an energy-dependent correction function,
which is approximately the fraction of energy going into electromagnetic
sub-showers \citep{AMZH}.
Originally calculated for cascades in ice, the
medium-dependence of this function has yet to be investigated, and so to
calculate the peak pulse strength for hadronic showers we used it 
unmodified with the electromagnetic result for the regolith described
above. The cone-width for hadronic showers was also taken from
\citet{AMZH}, extrapolated above 10 EeV as per \citet{williams}, and
scaled to the regolith as with the electromagnetic shower width.

In the Monte Carlo code the Cherenkov emission is represented as
bundles of ``rays'', each ray having associated with it a
direction, solid angle, field strength and polarisation.
Ray-tracing was used to propagate the radiation to Earth, at
distance $R=3.844 \times 10^8$-m. Modelled effects included the
electric-field attenuation length $\ell$ in the regolith ($\ell =
18$-m at 1 GHz; $\ell \propto 1/f$), refraction at the lunar
surface using the Fresnel transmission coefficients for each
component of the polarisation, and the solid-angle-stretching
factor applicable to a point source.

Surface roughness was simulated by randomly deviating the local
surface normal from the perpendicular. The deviation angle (the
adirectional slope) was calculated by generating the slope
tangents (unidirectional slopes) in orthogonal directions, which
are gaussian-distributed with mean 0 and variance $\arctan^2
(6^{\circ})$, where $\arctan (6^{\circ})$ is the RMS surface
roughness used in \citet{beresnyak}. This accounted for large-
scale surface roughness, such as hill-sides and crater walls,
with dimensions larger than the shower size. The effects of
intermediate-scale surface roughness, on scales between the
wavelength and shower size, is not currently understood
sufficiently to be included.

As with previous simulations, dispersion in the ionosphere was
assumed negligible within each 100 MHz sub-band, and the height
of the pulse was calculated by summing the contributions across
the bandwidth. The signal was assumed to retain its 100\%
polarisation, and thus be received with 50\% power efficiency by each of
the circularly-polarised receivers. As lunar thermal emission was
the dominant source of noise, the ratio of total signal (true UHE
neutrino signature plus a random noise component) to mean thermal
noise is preserved by the electronics; hence, detection could be
determined without the need to simulate the response of the
Parkes receiver system. This simplification is justified by the low 
background noise recorded when pointing off the Moon.

\section{Results} 
\label{results}
The simulation was run for energies in the range $10^{19}$-$10^{24}$
eV.  The calculated aperture for both centre-pointing and limb-
pointing configurations of the Parkes experiment is plotted in
Fig.~1, together with the simulation results for the GLUE experiment.
No aperture has been published for the Kalyazin experiment which had
a threshold of 13,500-Jy.  However, the simulation results of
\citet{beresnyak}, in which an otherwise identical experiment with
an assumed threshold of 3,000-Jy is modelled, have been included,
together with estimates for the improved Parkes experiment described
in Sec.~\ref{experiment}.

In comparing the apertures, note that the parameters we used to
simulate the Parkes experiment, such as depth of regolith,
amplitude of Cherenkov radiation, and mean surface roughness,
are at least as pessimistic as those of either
\citet{williams} or \citet{beresnyak}. Furthermore, perhaps because of parameter
differences, these two previous simulations produce apertures which
differ by an order of magnitude at $10^{23}$ eV, as first noted in
\citet{kalyazin}. This is surprising, firstly because of the
similarity of the modelled experiments, as opposed to the actual
experiments, and secondly because the greater difference is at high
energies where naively one would expect the results to be less
sensitive to differences in the modelling.

In more detail, consider lowering the simulated signal strength by 10\%.
This will produce a much lower effective aperture to neutrinos near the
threshold energy for neutrino detection, where all simulated detections
are marginal. However, a neutrino of $100$ times this energy will produce
a coherent Cherenkov signal with $100^2$ times the power, so that simulated
detections include only a very small fraction of marginal events. Hence,
the effect of lowering the modelled signal strength will be relatively
smaller for higher energy neutrino events. This is why the calculated
apertures have a stronger energy dependence at low neutrino energies: the
Parkes limb aperture at $10^{22}$ eV is 13.8 times that at $10^{21}$ eV,
but the aperture to $10^{24}$ eV neutrinos is only 3.9 times that at
$10^{23}$ eV (noting that a factor of 2.3 per decade arises naturally from
the increasing neutrino cross-section).

\begin{figure*}
\label{apertures}
{\psfig{file=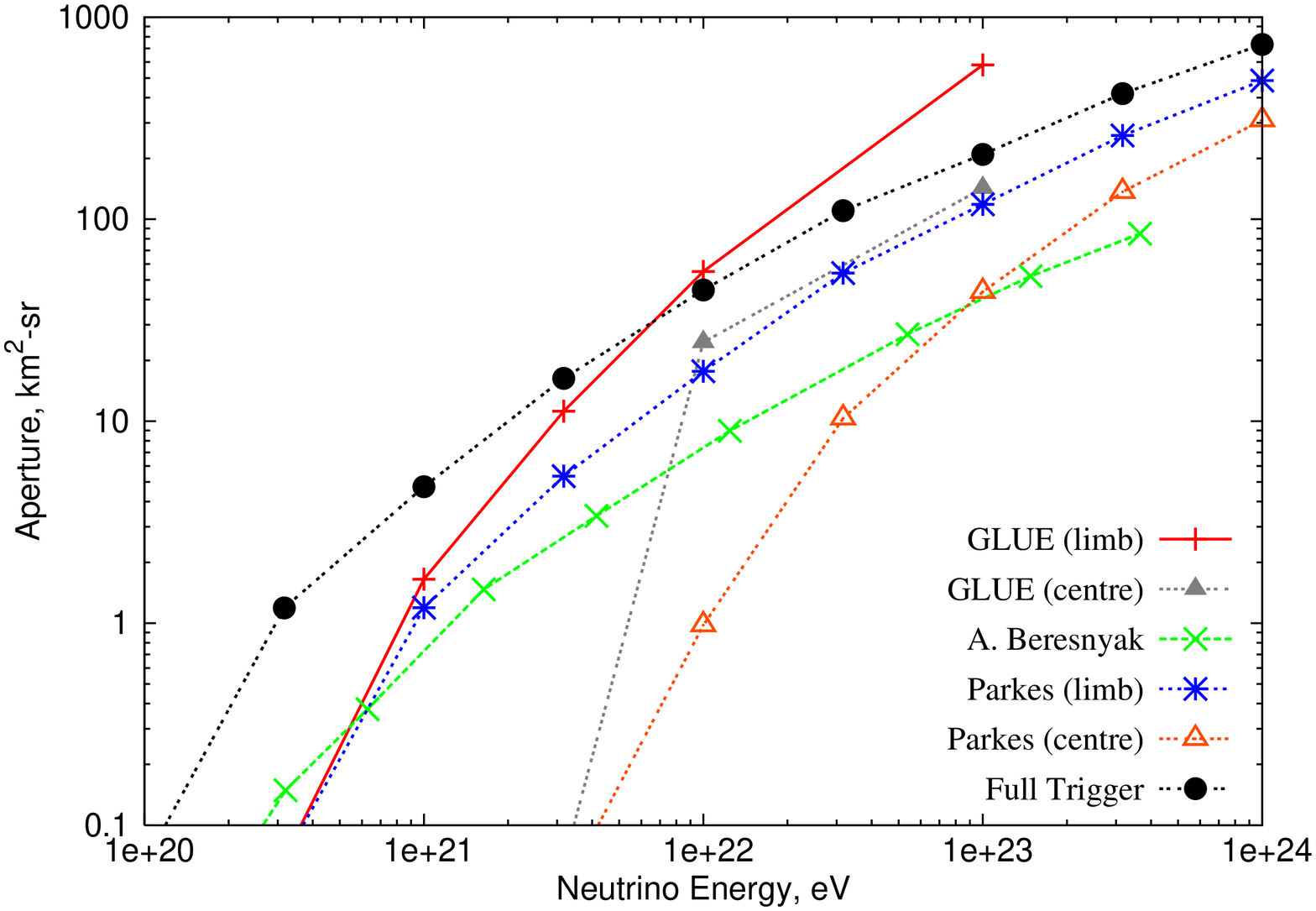, width=8cm, angle=270}}
\caption{(Colour online) Effective apertures (km$^2$-sr) as a function of
neutrino energy (eV) of Lunar radio-Cherenkov experiments: GLUE (from \citet{williams}), Kalyazin (from \citet{beresnyak}), and Parkes (our calculations). Also plotted are our estimates for Parkes in limb-pointing configuration had all available data been utilised in forming a trigger (`Full Trigger'). As discussed in the text, the plotted prediction for Kalyazin used an optimistic detection threshold, and the true sensitivity, particularly for the lower neutrino energies, will have been less.}
\end{figure*}

Putting these concerns aside, Fig.~1 clearly demonstrates
the benefits of the wide-bandwidth system used at Parkes. The 
effective energy threshold of $\sim 3 \times 10^{20}$ eV for the
limb-pointing configuration is similar to that achieved by both GLUE
(which utilised two antennae) and Kalyazin (basing this on the
simulation of \citet{beresnyak}, which assumed a sensitivity of 3,000-Jy instead of the eventual 13,500-Jy). Above threshold energies, the
Parkes aperture consistently lies below that from GLUE and above that
of Kalyazin, a comparison which could only be improved by the use of
identical simulation methods.

An unambiguous result is the desirability of pointing at the
limb. Compared to the limb-pointing, the
centre-pointing configuration of the Parkes experiment exhibits
an order of magnitude higher effective energy threshold ($\sim 4
\times 10^{21}$ vs $\sim 4 \times 10^{20}$ eV), and even at
$10^{23}$ eV the effective aperture is less than half that at the limb. This
effect was partially off-set in the GLUE experiment by
defocussing the 70-m dish when not pointing at the limb, which
accounts for the smaller difference in apertures between
configurations at $10^{22}$ eV. For experiments utilising lower
frequencies and smaller dishes, in which the FWHM of the beam is
comparable to, or greater than, the apparent size of the Moon,
the effect will likely be negligible.

The importance of developing signal processing techniques is shown by
the lower detection threshold and greater aperture which would have
resulted if the experiment at Parkes had been able to utilise all
available data in forming a trigger. As expected, the difference
is most pronounced at low energies, with the increased sensitivity
effected by the use of a wider bandwidth in triggering shifting the
effective aperture to the left. Methods to increase the aperture
at higher energies include the use of multiple beams to cover the
entire limb, and/or smaller antennas to cover the entire Moon with a
single beam. These will be discussed in a future paper

The effect of the lower observation frequency at Parkes is also
evident. As discussed in detail by \citet{scholten}, Cherenkov
radiation escapes the Moon more readily at low frequencies, and the
increase in beam size also allows more of the Moon to be observed.
The disadvantage is a lower sensitivity because the Cherenkov signal
is weaker at low frequencies, and the lower noise power per beam
solid angle from lunar thermal emission is mostly offset by the
increased beam size. The result is a steeper increase in aperture
with energy as is evident from Fig.~1  -- at $10^{21}$ eV, the
Parkes aperture is 1.5 times that of Beresnyak's result for Kalyazin,
whereas at $10^{23}$ eV it is three times larger.

The limit on the UHE neutrino flux we derive for the Parkes
experiment is plotted in Fig.~2.  The contribution
from the 8.5 hours spent pointing at the lunar centre is negligible
and, with only 2 hours of useful limb observations, the UHE
neutrino flux limit from the Parkes experiment is much weaker than that from either the
Kalyazin experiment (31.3 h \citep{kalyazin}) or Goldstone experiments (73.45 h
limb, 40 h half-limb \citep{williams}).

\begin{figure*}
\centerline{\psfig{file=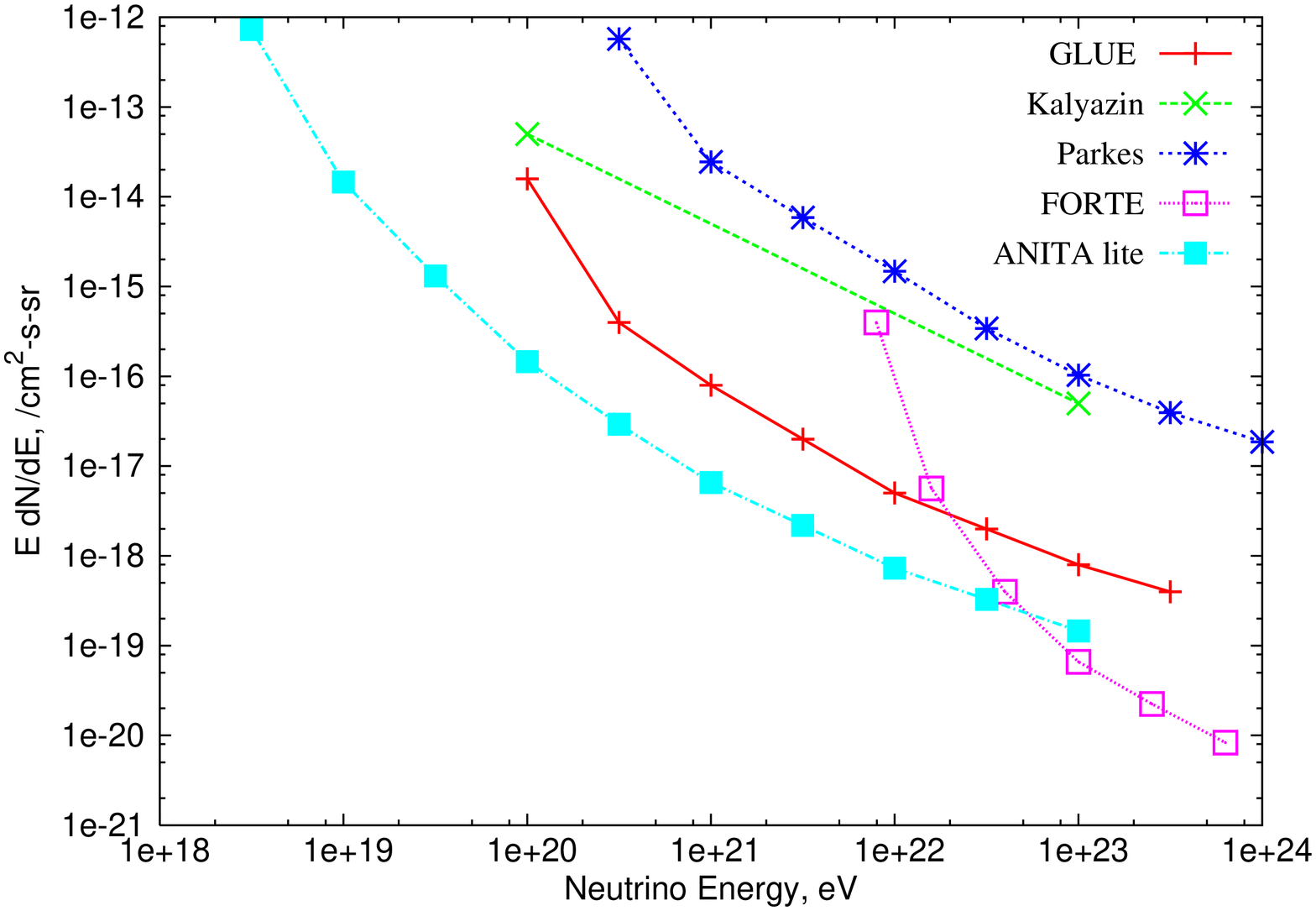, width=8cm, angle=270}}
\caption{(Colour online) UHE neutrino flux limits derived from lunar Cherenkov
observations. Our limit for Parkes (large dots), in accordance with limits for GLUE (from
\citet{williams}) and ANITA-lite (from \citet{anitalim}), was
calculated as a `model-independent' limit as per the prescription of \citet{forte}
for the FORTE limit. The limits from the Kalyazin experiment
are taken from \citet{kalyazin}, and were expressed as a 95\% confidence limit on an $E^{-2}$
spectrum between $10^{20}$ eV and $10^{23}$ eV.}
\label{limits}
\end{figure*}

\section{Conclusions}

We have calculated the effective aperture of the Parkes
experiment, and found that in the limb-pointing configuration it
was a competitive experiment as the exposure to UHE neutrinos per
hour of observation time was greater than the single-dish
experiment at Kalyazin, and it achieved an effective neutrino
detection energy threshold equal to that of the two-telescope
experiment at Goldstone.  Unfortunately, the centre-pointing
configuration, where most of the observation time was spent, had
negligible sensitivity at all but the highest energies. The
resulting limits on the UHE neutrino flux from the present Parkes
experiment are therefore not competitive with those from either
GLUE or Kalyazin.

Clearly, using the broad-band techniques at Parkes with modern
signal processing technology would greatly improve the sensitivity
of future searches for Lunar Cherenkov emission, even without
improved RFI discrimination. This could be provided by utilising
multiple antennas, as in GLUE, or by using intelligent hardware
which incorporates the discriminants currently used in off-line
processing into a real-time trigger. Future experiments should
aim to use both. The next generation of radio-telescopes, such as
LOFAR and the SKA, both of which will use large arrays of antennae
linked by high speed connections, will prove ideal instruments for
the UHE neutrino search, and may give the best chance to detect
these elusive particles.

\section*{Acknowledgments}
This research was supported under the Australian Research Council's Discovery funding scheme (project number DP0559991). Professor R. D. Ekers is the recipient of an Australian Research Council Federation Fellowship (project number FF0345330). C. W. James thanks Dr. T. Kneiske for assistance with simulation development, and J. Alvarez-Mu\~{n}iz for discussions on the scaling of coherent Cherenkov radiation.


\bsp
 
\label{lastpage}
 

\begin{thebibliography}{99}

\bibitem[\protect\citeauthoryear{Aggarwal \& Oberbeck}{1979}]{aggarwaloberbeck} Aggarwal H. R.,  Oberbeck V. R., 1979, Proc 10$^{\mathrm{th}}$ Lunar Planet. Sci. Conf., 2689

\bibitem[\protect\citeauthoryear{Alvarez-Mu\~{n}iz \& Zas}{1997a}]{AMZEM} Alvarez-Mu\~{n}iz J., Zas E., 1997, Phys. Lett. B, 411, 218 

\bibitem[\protect\citeauthoryear{Alvarez-Mu\~{n}iz \& Zas}{1997b}]{AMZ_scaling} Alvarez-Mu\~{n}iz J. \& Zas E., 1997, Proceedings of the XXV International Cosmic Ray Conference, Durban, South Africa, 7, 309

\bibitem[\protect\citeauthoryear{Alvarez-Mu\~{n}iz \& Zas}{1998}]{AMZH} Alvarez-Mu\~{n}iz J., Zas E., 1998, Phys. Lett. B, 434, 396

\bibitem[\protect\citeauthoryear{Alvarez-Mu\~{n}iz \& Zas}{2001}]{AMZmethods} Alvarez-Mu\~{n}iz J., Zas, E., 2001, astro-ph/0103369

\bibitem[\protect\citeauthoryear{Alvarez-Mu\~niz et al.}{2006}]{AMZlatest} Alvarez-Mu\~{n}iz et al., 2006, Phys. Rev. D, 74, 2, 023007

\bibitem[\protect\citeauthoryear{Askary'an}{1962}]{askaryan} Askary'an G. A., 1962, Sov. Phys. JETP, 14, 441

\bibitem[\protect\citeauthoryear{Barwick et al.}{2006}]{anitalim} Barwick S. \textit{et al}, 2006, Phys. Rev. Lett., 96, 171101

\bibitem[\protect\citeauthoryear{Beck}{2005}]{SKA} Beck, R.\ 2005, Astronomische Nachrichten, 326, 608

\bibitem[\protect\citeauthoryear{Beresnyak}{2004}]{beresnyak} Beresnyak A. R., 2004, astro-ph/0310295 v2

\bibitem[\protect\citeauthoryear{Beresnyak et al.}{2005}]{kalyazin} Beresnyak A. R. et al., 2005, Astronomy Reports, 49, 127


\bibitem[\protect\citeauthoryear{Billoir \& Bigas}{2006}]{auger} Billoir P., Bigas O. B., 2006, The Pierre Auger Observatory and neutrinos (Subm. to Neutrino Oscillation Workshop NOW 2006, September 9 - 16, 2006, Conca Specchiulla, Otranto, Italy)

\bibitem[\protect\citeauthoryear{Crocker et al.}{2005}]{flavourmixing} Crocker, R.~M.,
Melia, F., \& Volkas, R.~R.\, 2005, ApJ Lett.\, 622, L37

\bibitem[\protect\citeauthoryear{Dagkesamanskii \& Zheleznykh}{1989}]{dag} Dagkesamanskii R. D., Zheleznykh I. M., 1989, Sov. Phys. JETP Let., 50, 233

\bibitem[\protect\citeauthoryear{Falcke, Gorham \& Protheroe}{2004}]{prospects} Falcke H., Gorham P., Protheroe R. J., 2004, New Astron. Rev., 48, 1487

\bibitem[\protect\citeauthoryear{Gandhi et al.}{1998}]{gandhixsections} Gandhi R. et al., 1998,  Phys. Rev. D, 58, 093009

\bibitem[\protect\citeauthoryear{Gorham et al.}{2001}]{radhep} Gorham P. W. et al., 2001, in Saltzberg D., and Gorham P. W., eds, Radio Detection of High Energy Particles --- RADHEP 2000, AIP Conf. Proc. No. 579. AIP, New York, p177

\bibitem[\protect\citeauthoryear{Gorham et al.}{2004}]{goldstone} Gorham P. W. et al., 2004, Phys. Rev. Lett., 93, 041101

\bibitem[\protect\citeauthoryear{Gorham et al.}{2005}]{slac2} Gorham P. W. et al., 2005, Phys. Rev. D, 72, 023002

\bibitem[\protect\citeauthoryear{Gorham et al.}{2007}]{slac3} Gorham P. W. et al., 2007, hep-ex/0611008

\bibitem[\protect\citeauthoryear{Hankins, Ekers \& O'Sullivan}{1996}]{parkes} Hankins T. H., Ekers R. D., O'Sullivan J. D., 1996, MNRAS, 283, 1027

\bibitem[\protect\citeauthoryear{Hankins, Ekers \& O'Sullivan}{2001}]{parkesradhep} Hankins T. H., Ekers R. D., O'Sullivan J. D, 2001, in Saltzberg D., and Gorham P. W., eds, Radio Detection of High Energy Particles --- RADHEP 2000, AIP Conf. Proc. No. 579. AIP, New York, p168

\bibitem[\protect\citeauthoryear{Lehtinen \& Gorham}{2004}]{forte} 
Lehtinen, N. G., Gorham, P. W., 2004, Phys. Rev. D, 69, 013008

\bibitem[\protect\citeauthoryear{Mio\v{c}inovi\'c et al.}{2005}]{anita} Mio\v{c}inovi\'c P. et al., 2005, astro-ph/0503304 v1

\bibitem[\protect\citeauthoryear{Mio\v{c}inovi\'c et al.}{2006}]{time_domain} 
Mio\v{c}inovi\'c P. et al., 2006, Phys. Rev. D, 64, 043002


\bibitem[\protect\citeauthoryear{Saltzberg et al.}{2001}]{slac}  Saltzberg D. et al., 2001, Phys. Rev. Lett., 86, 2802


\bibitem[\protect\citeauthoryear{Scholten et al.}{2006}]{scholten} Scholten O. et al., 2006, Astropart. Phys., 26, 219


\bibitem[\protect\citeauthoryear{Shkuratov \& Bondarenko}{2001}]{shkuratov} Shkuratov Y. G., Bondarenko N. V., 2001, Icarus, 149, 329

\bibitem[\protect\citeauthoryear{Short \& Forman}{1972}]{short} Short N. M., Forman M. L., 1972, Modern Geology, 3, 69



\bibitem[\protect\citeauthoryear{Seckel \& Stanev}{2005}]{stanev} Seckel D., Stanev T., 2005, Phys. Rev. Lett. D, 95, 041101

\bibitem[\protect\citeauthoryear{Williams}{2004}]{williams} Williams D. R., 2004, The Askar'yan Effect and Detection of Extremely High Energy Neutrinos in the Lunar Regolith and Salt, Dissertation, University of California


\end{thebibliography}
\end{document}